\documentclass[12pt]{article}
\usepackage{hyperref}
\usepackage{graphicx}
\usepackage[hang,small,bf]{caption}
\newcommand{\be}[3]{\begin{equation}  \label{#1#2#3}}

\newcommand{\ee}{ \end{equation}}
\newcommand{\ba}{\begin{array}}
\newcommand{\ea}{\end{array}}

\let\Large=\large

\begin{document}
\thispagestyle{empty}
\begin{flushright}
CALT-68-2230 \\
hep-th/9907070  
\end{flushright}
\vspace{2cm}

\begin{center}
\bf \Large 
Domain walls of D=5 supergravity and \\[2mm]
fixed points of $N$=1 Super Yang Mills
\end{center}

\vspace{2cm}

\centerline{{\bf Klaus Behrndt}\footnote{Email: behrndt@theory.caltech.edu}
}

\bigskip

\centerline{\em California Institute of Technology}

\centerline{\em Pasadena, CA 91125, USA}

\vspace{1cm}

\begin{abstract}
\noindent
Employing the AdS/CFT correspondence, we give an explicit supergravity
picture for the renormalization group flow of couplings 4-d super Yang
Mills with four supercharges. The solution represents a domain wall of 5-d,
N=2 supergravity, that interpolates between two (different) $AdS_5$ vacua
and is obtained by gauging a $U(1)$ subgroup of the $SU(2)$ R-symmetry.
On the supergravity side the domain wall couples only to scalar fields from
vector mulitplets, but not to scalars from hyper multiplets. We
discuss the $c$-theorem, the $\beta$-functions and consider two examples:
one is the sugra solution related to $Z_k$ orbifolds (corresponding to
N=2 SYM) and the other is an orientifold construction for an elliptically
fibered CY with $F_1$ basis (corresponding to N=1 SYM).
\end{abstract}

\vspace{5mm}

\newpage


\section{Introduction}


Many field theories flow under the renormalization group (RG) towards
fixed points, where they become finite and universal (scheme independent).
However, the fixed point values of the couplings are not necessarily small
and generically perturbation theory breaks down. On the other hand,
employing the AdS/CFT correspondence \cite{330, 340, 320} we can
formulate these interacting field theories in terms of AdS gravity. Super
Yang-Mills in 4 dimensions e.g., becomes conformal near fixed points and is
expected to be dual to $AdS_5$ gravity -- at least in appropriate limits.
The flow between two different fixed points corresponds in supergravity to
domain walls (DW), which are kink solutions connecting two vacua which
appear as extrema of the potential (typically AdS spaces). For a
review on domain walls see \cite{180} and different aspects of the
AdS/CFT correspondence (or more general DW/QFT correspondence) are
discussed in \cite{480, 440}.

Domain walls are supported by at least one scalar field and appear
naturally in supergravity where a subgroup of the R-symmetry has been
gauged. As a consequence of the gauging the sugra Lagrangian contains a
potential and the flow of the field theory is encoded in non-trivial
supergravity scalars that run from one extremum of the potential to
another. {From} these fixed scalars it is straightforward to obtain
the corresponding fixed couplings in the field theory. If the potential
allows only for one extremum the field theory has either
an IR or UV fixed point, but not both.

For a concrete supergravity potential, it is straightforward to expand
a solution around a given fixed point. However, in order to get a global
picture, we need an explicit solution that interpolates between two
(different) $AdS_5$ vacua on both sides of the wall. Having this
solution it should be straightforward to calculate field theory
quantities like $\beta$-functions or anomalous dimensions.

The qualitative picture depends very much on the amount of unbroken
supersymmetry and only in field theory with at most four supercharges we
can expect a flow between two non-trivial fixed points.  One way to break
partly supersymmetry has been discussed in \cite{072, 070, 060} and
represents a non-abelian gauging of D=5, N=8 supergravity.

Another possibility is to gauge a generic N=2 supergravity which has
in the ungauged case only eight supercharges and the domain wall will
break at least one half of them. In these models the scalar fields
enter two different multiplets: the vector multiplets or
hypermultiplets, and for many purposes it is reasonable to consider one
sector only. This truncation is always possible as long as the scalars
are not charged.
An interesting domain wall solution for this
non-abelian gauging has been discussed in \cite{260} and it contains
an arbitrary number of vector multiplets which couple to the universal
hypermultiplet; this formulation has been extended to include
non-universal hypermultiplets \cite{350}.

In this paper we will explore the situation where only a $U(1)$
subgroup of the $R$-symmetry has been gauged and the hypermultiplets
remain uncharged and we can consistently decouple this sector. As we
will see, these domain walls provide a supergravity solution for a
non-trivial RG flow connecting two fixed points. Since this solution
holds for any prepotential, we obtain an exact expression in the
two-derivative approximation for the $\beta$-functions.

We have organized our paper as follows. We start with a discussion of
the domain wall solution and show that it represents a BPS
configuration. In section 3 we will describe the sypersymmetric flow
between extrema of the superpotential and we formulate the
$c$-theorem, where the $C$-function will determine the
$\beta$-functions. The domain wall solution is quite general and is
not related to a specific example or field theory. To be concrete we
discuss in section 4 two examples, one related to N=2 and the other to
N=1 super Yang Mills, where the scalars run from the boundary of the
moduli space towards an $SU(2)$ enhancement point. The central charges
and $\beta$-functions are calculated. In the conclusion we will
summarize our results.


\section{Abelian gauged supergravity in 5 dimensions}


\subsection{General remarks}


Let us start with some general remarks about gauged $N$=2 supergravity in
5 dimensions, see \cite{020, 300} for more details. As mentioned in the
introduction we are interested in an abelian gauging and can truncate the
model to the vector multiplets sector only. The bosonic Lagrangian is
then given by
\be100
S_5 = \int \Big[{\frac{1}{2}} R + g^2 P - 
{\frac{1}{4}} G_{IJ} F_{\mu\nu}
{}^I F^{\mu\nu J}-{\frac{1}{2}} g_{AB} \partial_{\mu} 
\Phi^A \partial^\mu \Phi^B \Big] +
{1 \over 48} \int C_{IJK}
F^I \wedge F^J \wedge A^K  
\ee
where $(\mu,\nu)=0,1,\cdots, 4$ are space-time indices; $A,B = 1,
\ldots , n$ count the number of vector multiplets (each containing
one real scalar $\Phi^A$ and one gauge field), in addition there is one
gauge field in the gravity multiplet so that $I,J = 0, \ldots n$. Both
the gauge fields and the scalar part have non-trivial couplings $G_{IJ}$
and $g_{AB}$ which depends on the scalar fields.  The potential $P$
arises due to the gauging of an $U(1)$ subgroup of the $R$-symmetry and
the corresponding gauge field is a linear combination of all other
abelian gauge fields
\be101
A_\mu=V_I A_\mu^I 
\ee
where $V_I$ are constants (Fayet-Illiopoulos terms). Notice, by this
abelian gauging fermions become charged, but all scalars from the vector
as well as hypermultiplets remain neutral. In contrast, under a non-abelian
gauging the hypermultiplets become charged and one can not ignore them.

We are especially interested in a solution, which describes 
a 3-brane or domain wall living in a 5-d asymptotic anti-de
Sitter space. We assume that it is flat, static and isotropic
which yields as ansatz for the metric
\be102
ds^2 = e^{2U} g^2 r^2 \Big[ -dt^2 + dy_1^2 + dy_2^2 + dy_3^2 \Big]
	+ e^{-4U} {dr^2 \over g^2 r^2} \ ,
\ee
where $U$ is a function of the radius $r$ and becomes constant
in the asymptotic AdS vacuum.

The gauging can be viewed in different ways. {From} the M- or F-theory
perspective, one compactifies on a Calabi-Yau, but with {\em
non-trivial} (=non-zero mode) internal components of the 4-form field
strength. As discussed in \cite{390} this is required for a consistent
reduction of the Horava-Witten model \cite{400} and yields naturally
3-branes in 5 dimensions. On the other hand from the 10-d perspective
\cite{290, 380}, the space transverse to the 3-branes is given by a
cone, $ds_{\perp} \sim dr^2 + r^2 ds_X^2$, over the horizon manifold
$ds_X^2$ which is generically not spherical symmetric. This is the
case at least as long as one is in a region which we call $AdS_5$
vacuum, where the scalars become constant and which correspond to
fixed points. In general however, one cannot expect that the 10-d metric
factorizes, e.g., non-extreme AdS black holes can be interpreted as
rotating 3-branes \cite{410, 490, 540}. To get a better understanding
of our solution it may help to adopt the M- or even F-theory
approach.

Before we come to the generic N=2 case,
we will discuss the solution that can naturally be embedded into $N$=4,8
supergravity.


\subsection{A simple example}


In the past years we have learned a lot about 5-d black holes and
string-type solutions, which are the natural objects that are charged.
Much less is known about domain wall solutions. The main difference is
that black holes are well defined in an asymptotic flat spacetime whereas
domain walls are asymptotically AdS, if we impose that the asymptotic
space is maximal supersymmetric. But it is straightforward to truncate,
or if one likes to promote, any (static) black hole into a domain wall.
What one has to do is:

(i) embed it into an AdS space, \newline 
(ii) replace the $S_3$-horizon by $S_{3,k}$ 
(3-sphere with constant curvature $k$) 
and \newline 
(iii) take the limit $k \rightarrow 0$.

The first step has been discussed in \cite{010}, the second one in \cite{030} 
and the last one is trivial, see also \cite{120, 410, 490, 540, 550}
where other dimensions have also been discussed.

Let us explain it for the simple example of the 3-charge solution.
This is an interesting example, because due to the solutions
generating technique \cite{050, 250}, any black hole that can be
embedded into ungauged $N$=4,8 supergravity can be obtained {from}
this solution\footnote{However the $U$-duality group is broken by
the gauging and we will not discuss here to which extend the
solution generating technique is applicable.}. 

After step (i) and (ii) the extreme solution reads \cite{030}
\be090
\ba{l}
ds^2 = - {  f \over (H_1 H_2 H_3)^{2/3}}\,  dt^2 + (H_1 H_2 H_3)^{1/3} 
   \Big[ {dr^2 \over f} + r^2 d\Omega_{3,k} \Big] 
   \ , \ A^I_0 = \sqrt{k}/ H_I   \ , \\
	\Phi^1 = {(H_1 H_2 H_3)^{1/3} \over H_1} \quad , \quad 
	\Phi^2 = {(H_1 H_2 H_3)^{1/3} \over H_2 } \quad , \quad  
	 f = k + g^2 r^2 H_1 H_2 H_3
\ea
\ee
where for $k=1$ the horizon is a sphere and for $k=-1$ it becomes a
hyperboloid. In the limit $k \rightarrow 0$ this solution becomes a
3-brane compatible with our ansatz (\ref{102})
\be110
\ba{l}
ds^2 = (H_1 H_2 H_3)^{1/3} g^2 r^2 \Big[ -dt^2 +dy^2_1 + dy_2^2
	+dy_3^2 \Big] + {dr^2 \over g^2 r^2 (H_1 H_2 H_3)^{2/3}} \ ,\\
	\Phi^1 = {(H_1 H_2 H_3)^{1/3} \over H_1} \quad , \quad 
	\Phi^2 = {(H_1 H_2 H_3)^{1/3} \over H_2 } \quad , \quad  
	F^I_{\mu\nu} =0  
\ea
\ee
where the harmonic functions $H_I$ are given by
\be120
H_{1,2,3} = 1 + {q_{1,2,3} \over r^2}  \ .
\ee
For this case the supergravity potential is (see below)
\be130
P = 2 \Big( {1 \over \Phi^1} + {1 \over \Phi^2} + \Phi^1 \Phi^2 \Big) \ ,
\ee
with the minimum at $\Phi^1 = \Phi^2 = 1$. Let us stress, that in order
to perform the truncation to a flat ($k=0$) 3-brane, it is crucial to
consider gauged supergravity with an asymptotic AdS vacuum -- the extreme
ungauged solution (with no potential) cannot be truncated to a flat
3-brane! Let us also mention, that in the non-extreme case the
gauge fields survive the domain wall limit $k \rightarrow 0$,
see \cite{410, 490, 540}.

This solution contains three classes corresponding to: one, two or
three non-trivial harmonic functions. Asymptotically all solutions
become $AdS_5$, but they differ near the core. To discuss the different
cases in more detail we may equalize all non-vanishing harmonics,
which means that we replace $H_1 H_2 H_3 = H^{n}$ ($n = 1,2,3$).
Choosing a coordinate system where the metric becomes
\be140
ds^2 = e^{2A} \Big[ -dt^2 + dy_1^2 + dy_2^2 + dy_3^2 \Big] + d\rho^2 
\ee
we find near $r \simeq 0$
\be150
e^{2A} \sim \left\{ \ba{ll} q \, g^4 (\rho -a)^2 \ , & n = 1 \\
	q \, g^{5/2} \sqrt{|\rho -a|}\ , \qquad & 	n = 2 \\
	q \, g^2 \ ,	& n =3 \ . \ea \right.
\ee
where $a$ is an arbitrary parameter. Only the case $n=3$ behaves
smooth near $\rho=a$ (or at $r=0$), the $n=2$ case has a curvature
singularity and for $n=1$ the metric exhibits a conical singularity
(the case $n=2$ has been addressed also in \cite{110, 460}) In
addition, after equalizing the harmonic functions we can have at most
one scalar field, which either vanishes or diverges near the origin
and plays the role of the dilaton. This singularity indicates that for
these two cases the corresponding Yang-Mills coupling runs either to a
strongly or weakly coupled regime (electric or magnetic picture cp.\
\cite{040}).

Perhaps more interesting than these singular cases is the case $n=3$
where the spacetime is regular at $r=0$ ($\rho = a$). By equalizing
all three harmonics all scalars are trivial and the metric reads 
\be155
\ba{rcl}
ds^2 &=& H g^2 r^2 \Big[ -dt^2 +dy^2_1 + dy_2^2 +dy_3^2 \Big] + 
      {dr^2 \over g^2 r^2 H^2} \\
      &=& g^2 \rho^2 \Big[ -dt^2 +dy^2_1 + dy_2^2 +dy_3^2 \Big] + 
      {d\rho ^2 \over g^2 \rho^2} \ ,
\ea
\ee
which is nothing other than $AdS_5$ ($r^2 H \equiv r^2 + q =
\rho^2$). So, in order to obtain a non-trivial solution we need
different $H's$, or equivalently, we have to turn on scalar
fields. But still, the metric will remain regular at $r=0$ where $\rho
= \sqrt{q}$ and spacetime does not end there. This is very 
similar to the extreme Reissner-Norstr\"om solution, but with the
difference that we do not have a horizon in the case at hand. In fact
near $r=0$ the metric (\ref{110}) becomes
\be160
ds^2 = (q_1 q_2 q_3)^{1/3} g^2 \Big[ -dt^2 + dy^2_1 + dy_2^2
	+dy_3^2 \Big] + {r^2 dr^2 \over g^2 (q_1 q_2 q_3)^{2/3}} \ .
\ee
Since this metric, as well as the scalars, are invariant under the $Z_2$
symmetry: $r\rightarrow -r$ we can continue the solution beyond the point
$r=0$. If we do not break this reflection symmetry, we effectively
identify the two asymptotic regions, which is equivalent to a compact
transverse space with a radius given by the cosmological constant. Note,
the infinite radial part of an AdS space can always be mapped on a finite
interval. This case corresponds to the situation described in
\cite{130, 270}, where the radial coordinate is
an angle and the identification $r \simeq -r$ yields an $S/Z_2$
orbifold as transversal space.  For earlier work on reflection
symmetric domain walls see \cite{530}.  But a generic domain wall
breaks the discrete symmetry and thus we may treat both sides
differently. One possibility would be to connect the solution to flat
space, which means that the potential has to vanish. This
supersymmetry breaking setup has been explored in \cite{370, 310}.

In summary, the 3-brane as given in (\ref{110}) represents a domain
wall interpolating between two $AdS$ spaces or an $AdS$ space and flat
spacetime.  In order to match the scalars and the metric, we have to
equalize the $q's$ on both side and if one wants to describe a flat
space on one side, the constant parts of the harmonics have to vanish
on that side (see also below).  When discussing genuine $N$=2
solutions in the next section, we will see that we can glue together
topological distinct vacua at this point, where each of them may have
a different cosmological constant. In the dual field theory this setup
will correspond to the incorporation of additional perturbations and
we expect a flow between different fixed points.


\subsection{Definitions and conventions of 5-d supergravity}


A generic solution has of course more scalars and is in the ungauged
case not duality equivalent to a 3-charge solution, which we discussed
in the last section. The $n$ physical scalars $\Phi^A$ of 5-d
supergravity parameterize a hypersurface in an $(n+1)$-dimensional
space parameterized by the coordinates $X^I$. This space will be
called scalar manifold or occasionally moduli space, but we have to
keep in mind that the gauging breaks the $U$-duality group.
Let us summarize some basic features; we will
mainly follow here \cite{360, 310, 020}, but in slightly modified
notation. To be concrete the scalar manifold is defined by the constraint
\be180
{\cal V} = {1 \over 6} C_{IJK} X^I X^J X^K = 1   \ ,
\ee
where in the case of a Calabi-Yau compactification the constants
$C_{IJK}$ are the topological intersection numbers. For many
physical interesting cases this space is given by a coset like \cite{020}
\be182
{\cal M} = {SO(n-1,1) \over SO(n-1)} \times SO(1,1) \ .
\ee
In general there is no restriction in the number of scalar fields, the
case discussed in the last section corresponds to $n=2$ and
$C_{123} =1$.

The coupling matrices entering the Lagrangian (\ref{100}) are 
defined by
\be190
G_{IJ} = - {1 \over 2} \Big(\partial_{I} \partial_J \, \log {\cal V}
\Big)_{{\cal V} = 1}
\qquad , \qquad
g_{AB} = \Big(\partial_A X^I \partial_B X^J G_{IJ} \Big)_{{\cal V}=1} 
\ee
where $\partial_A \equiv {\partial \over \partial \Phi^A}$ and
$\partial_I \equiv {\partial \over \partial X^I}$. It follows that
\be192
\ba{l}
G_{IJ} = - {1 \over 2} \Big[ C_{IJ} - {1 \over 4} C_I C_J \Big] \\
\partial_K G_{IJ} = -{1 \over 2}
\Big[ C_{IJK} - {1\over 2}(C_{IJ} C_K + {\rm cycl.}) +
{1 \over 4} C_I C_J C_K \Big] \ ,
\ea
\ee
with $C_I = C_{IJK} X^J X^K$ and $C_{IJ} = C_{IJK} X^K$. Moreover,
the normal vector on the scalar manifold is given by\footnote{In
proper coordinates, the defining equation of a surface $F(X^I)=1$
depends only on the transversal coordinates, at least locally.
Therefore, $\partial_I F$ is a normal vector and $\partial_A F 
\equiv (\partial_A X^I) \partial_I F(X) = 0$.}
\be191
X_I \equiv {2 \over 3} G_{IJ} X^J = {1 \over 6} C_{IJK} X^J X^K = {1
\over 3} \partial_I {\cal V}
\ee
(normalized as $X_I X^I = 1$). It follows that
\be194
\partial_A X_I = - {2 \over 3} G_{IJ} \partial_A X^J \ ,
\ee
and since $\partial_A X^I$ gives the tangent vectors
\be193
X_I \partial_A X^I = - X^I \partial_A X_I = 0 \ .
\ee
Finally, the potential $P$ in the Lagrangian (\ref{100}) reads
\be218
\ba{rcl}
P &=& 6\, V_I V_J \Big( X^I X^J - {3 \over 4} 
g^{AB} \partial_A X^I \partial_B X^J \Big) \\
&=& 6\, \Big( W^2 - {3 \over 4} g^{AB} \partial_A W \partial_B W \Big) \ ,
\ea
\ee
where $V_I$ are constants (FI terms) and the superpotential $W$ is
\be216
W = V_I X^I \ .
\ee
Notice, $W$ is subject to the constraint (\ref{180}) which makes it
non-linear in the physical scalars $\Phi^A$.


\subsection{BPS domain walls in gauged $N$=2 supergravity}


We are interested in a 3-brane, that couples to $n$ scalars of the vector
multiplets and for which the gauge fields are trivial.  This domain wall
solution allows for unbroken supersymmetries if the gaugino and gravitino
variations\footnote{In our case of abelian gauging, the hyperrino
variation are trivially solved for constant hyper scalars.}
\be210
\ba{l}
\delta \lambda_A = \Big( - {i \over 2} g_{AB} \Gamma^{\mu}
\partial_{\mu} \Phi^B + i\, {3 \over 2} g  \partial_A W \Big)
\epsilon \ ,\\
\delta \psi_{\mu} = \Big( \partial_{\mu} + 
{1 \over 4} \omega_{\mu}^{ab} \Gamma_{ab}
+ {1 \over 2} g \, \Gamma_{\mu} W \Big) \epsilon \ ,
\ea
\ee
have non-trivial zeros. Solutions of these equations are BPS
configurations of $N$=2 gauged supergravity and basically there are two
dual cases related to the possibilities to express the scalar fields in
terms of $X_I$ or its dual $X^I$. Both coordinates parameterize dual
cycles of the internal manifold, $X_I$ is related to 4-cycles whereas
$X^I$ to 2-cycles. Taking into account non-trivial gauge fields, the
solution expressed by $X_I$ corresponds to electric  whereas
the other to magnetic solutions. In this paper we will
explore only the electric solution. Most likely the magnetic solution is
not supersymmetric, at least as long as one keeps the ansatz for $U(1)$
gauge fields $A_{\mu} = V_I A^I_{\mu}$, where $V_I$ is a constant
``electric'' vector which by supersymmetry will be related to the electric
moduli (see below). For a discussion of electric-magnetic duality
in gauged supergravity see \cite{370}.

Our (electric) solution reads
\be200
\ba{l}
ds^2 = e^{2U} g^2 r^2 \Big[ -dt^2 +dy^2_1 + dy_2^2
	+dy_3^2 \Big] + e^{-4U} {dr^2 \over g^2 r^2} \ ,\\
	X_I = {1 \over 3} e^{-2U} H_I \qquad , \qquad  H_I = h_I
	+ { q_I \over r^2} \\
\ea
\ee
where all gauge fields vanish.  As it has been shown in \cite{030}
this configuration solves the equations of motion coming from the
Lagrangian (\ref{100}) for any prepotential. What remains
to be shown is that this domain wall allows for unbroken
supersymmetry. 

Let us start with the gaugino variation. Using the definition
(\ref{190}) and (\ref{194}) we find
\be220
g_{AB} \partial_{\mu} \Phi^B = G_{IJ} \partial_A X^I \partial_{\mu} X^J
= - {3 \over 2} \partial_A X^I \partial_{\mu} X_I
\ee
and therefore the gaugino variation becomes
\be230
\delta \lambda_A = {3\,i \over 4} \partial_A X^I \Big(\Gamma^{\mu}
\partial_{\mu} X_I + 2 g V_I \Big) \epsilon \ .
\ee
It follows from (\ref{193}) and our solution (\ref{200}) that
\be240
(\partial_A X^I) X_I = {1 \over 3}\, e^{-2U} (\partial_A X^I) H_I = 0 
\ee
and thus
\be250
 \partial_A X^I \Gamma^{\mu} \partial_{\mu} X_I = 
 e^{-2U} \, {2 \over 3 r} \, 
\partial_A X^I h_I \, \Gamma^{\underline r}\ .
\ee
After transforming $\Gamma^{\underline r}$ into the tangent space,
the gaugino variation becomes
\be260
\delta \lambda_i \sim \partial_A X^I \Big( \Gamma^r h_I + 3 V_I \Big)
\epsilon \ .
\ee
Taking the projector
\be270
(1 + \Gamma^r)  \epsilon = 0 
\ee
the constants in the harmonic functions are fixed by the $V_I$ 
vector
\be272
h_I =  3 V_I \ .
\ee
Notice, treating $\Gamma^r$ as $\Gamma^5$, we project out one
chirality from the 4-d perspective.  The chirality discussed in this
paper yields a Killing spinor (see eq.\ (\ref{330})) which is located
near $r=0$; whereas the Killing spinor with respect to the other
chirality would be located at $r= \infty$; the
location is given by an exponentially increase or
fall-off in adapted coordinates.

Taking the normalization that $e^{2U} \rightarrow 1 $ for $r\rightarrow
+ \infty$, we find that 
\be273
X_I \big|_{+ \infty} =  V_I
\ee
i.e., the moduli fix the constant vector $V_I$.

Next we turn to the gravitino variation and find for the non-vanishing
spin connections
\be280
\omega^{0r} = g^2 r e^{2U} (r e^U)' \, dt \qquad , \qquad
\omega^{mr} = g^2 r e^{2U} (r e^U)' \, dx^m \ .
\ee
As a consequence of the scalar constraint $X_I X^I =1$, $e^{2U}$ 
becomes
\be290
e^{2U} = {1 \over 3} X^I H_I
\ee
and therefore (recall $H_I \partial X^I = 0$)
\be300
(e^{2U})' = {2 \over r} \Big({1 \over 3} X^I h_I - e^{2U} \Big) \ ,
\ee
or
\be302
\Big( r e^{U} \Big)' = {1 \over 3}  (X^I h_I) \, e^{-U} = W \, e^{-U} \ ,
\ee
with $W$ defined in (\ref{216}). Using this relation we find that the
worldvolume components of the gravitino variation
\be310
\Gamma_{\alpha}\, \Big[ {1\over 2} g^2 r e^{2U} (r e^U)' \,  \Gamma_r + 
{1 \over 2} g^2 \, r e^U W \Big] \epsilon 
= 0
\ee
vanish ($\alpha = 0 \ldots 3$). In order to determine the Killing
spinor $\epsilon$ we have to solve $\delta \psi_r =0$, which becomes
\be320
\delta \psi_r = \Big[ \partial_r + 
{1 \over 2 r} e^{-2U} W \, \Gamma_r \Big] \epsilon =
\Big[ \partial_r + {1 \over 2}\partial_r \Big( U + \log r \Big) \, 
\Gamma_r \Big] \epsilon = 
0 \ .
\ee
Using the projection  (\ref{270}) it is straightforward to solve
this differential equation and the Killing spinor reads
\be330
\epsilon = e^{- {1\over 2} (U + \log r)} 
\Big( 1 - \Gamma_r \Big) \epsilon_0
\ee
where $\epsilon_0$ is an arbitrary constant spinor.

This completes the discussion of supersymmetry.  We have shown that 
the solution (\ref{200}) represents a BPS domain wall of gauged $N$=2
supergravity.

One may of course ask whether this is just a special solution or to
which extent it is general. Let us add some comments. First, that the
projector (\ref{270}) depends only on $\Gamma^r$ is dictated by the
geometry (isotropic 3-brane) and employing eq.\ (\ref{193}) the
gaugino variation becomes
\be340
\partial_A X^I \, ( - r e^{2U} \partial_{r} X_I + 2  V_I ) \epsilon 
\equiv \partial_A X^I \, \Big( - r  \partial_{r}(e^{2U} X_I) + 
2  V_I \Big) \epsilon = 0 \ .
\ee
with the solution given by
\be342
 - r \partial_{r}(e^{2U} X_I) + 2 V_I = A X_I
\ee
for some function $A$. To solve the equation
explicitly we contract it with $X^I$ and get: $r \partial_r e^{2U}
- 2 W = -A$. In order to fix $A$ we use the gravitino variation, which
yields
\be343
W = e^U \partial_r \big( r e^U \big)
\ee
and hence $A = 2 \, e^{2U}$. Inserting this back in (\ref{342}) we get
the unique solution
\be350
e^{2U} X_I = {1 \over 3} \, \Big(  3 V_I + {q_I \over r^2}\Big) 
= {1 \over 3} \, H_I \ ,
\ee
(the ${1 \over 3}$ is a convenient normalization). Thus, eq.\
(\ref{200}) is not only a special ansatz, it represents the generic
isotropic 3-brane solution of 5-dimensional gauged supergravity with
trivial hypermultiplets and AdS boundary condition. These equations are
also known as stabilization equations and have been first discussed for
the black hole entropy \cite{160} and later for general stationary
BPS solutions N=2 sugra in \cite{420, 430}.


\section{Supersymmetric fixed points, RG flow and the $c$-theorem}


The AdS/CFT correspondence tells us that the renormalization group flow
of Yang-Mills couplings translate into a running (radial dependent) 
scalar fields. In this setup fixed points correspond to regions where
the scalars become constant and extremize the potential; the metric is
then anti de Sitter. In this section we will discuss some general
features of this flow and afterwards we will present examples.

Let us start by rewriting the metric in a different way.
Defining
\be400
r\, e^{U} = \mu
\ee
and using (\ref{302}) we get $\mu d \mu = W rdr$ and thus
\be408
ds^2 = g^2 \mu^2 \Big( -dt^2 + dy_1^2 +dy_2^2 + dy_3^2 \Big)
+ { d\mu^2 \over ( g W )^2 \mu^2} \ .
\ee
Therefore, whenever we approach an extrema of the superpotential where
$\partial_A W = 0$ and thus $W$ becomes constant, we obtain an $AdS_5$
space with a radius given by $l = 1/(gW_0)$ and fixed scalars. In
fact, from the gaugino variation it follows, that if the scalars are
constant the superpotential is extremal
\be470
\partial_A W = 0 
\ee
which implies that also the supergravity potential is extremal
($\partial_A P =0$). In the dual Yang-Mills theory we have reached a
conformal fixed point. The scalars are generically ratios of harmonic
functions and therefore only $r = \infty$ or $r=0$ could correspond to
fixed points. But as we will see below, the point $r=0$ is not a fixed point
but a phase transition point.

But let us first expand the potential around a fixed point.  The first
derivative vanishes and the second derivative gives
\be480
\partial_A \partial_B W = {1 \over 3} h_I \partial_A \partial_B X^I \ .
\ee
{From} the definition of the scalar metric we get
\be490
g_{AB} = \partial_A X^I \partial_B X^J G_{IJ} = - {3 \over 2}
\partial_A X_I \partial_B X^I = - 
{3 \over 2} \partial_A \Big(  X_I \partial_B X^I \Big) + 
{3 \over 2}  X_I \partial_A \partial_B X^I 
\ee
where the first term on the rhs vanishes identically, see eq.\
(\ref{240}). Inverting this equation yields (recall $X_I X^I =1$) 
\be492
\partial_A \partial_B X^I = {2 \over 3} g_{AB} X^I + T_{AB}^{\ \ D}
\partial_D X^I
\ee
where $T_{ABC}$ is defined as the projection of $C_{IJK}$ on the scalar
manifold \cite{360}
\be494
T_{ABC} = \partial_A X^I \partial_B X^J \partial_C X^K C_{IJK} \ .
\ee
Contracting (\ref{492}) with $V_I$ yields
\be495
\partial_A \partial_B W = {2 \over 3}\, g_{AB} W + T_{AB}^{\ \ C} \partial_C W
\ee
and thus we get at the fixed point ($\partial_C W = 0$) the well-known relation
\be500
\partial_A \partial_B W \Big|_0 = {2 \over 3} g_{AB} W_0 \ .
\ee
Hence, supersymmetric fixed points (extrema) remain stable as long as the
scalar metric $g_{AB}$ is positive definite. Notice, it is $W$ that
drives the supersymmetric flow and {\em not} the sugra potential
$P$. The expansion can be continued and we find for the third
derivative near the extremum
\be502
\partial_A \partial_B \partial_C W\Big|_0 = {8 \over 3} \, W_0 \, T_{ABC}
\ .
\ee
Putting the terms together, the superpotential becomes
\be503
W = W_0 \, \Big( 1 + {1 \over 3} \, g_{AB} \phi^A \phi^B + {4 \over 9}
\, T_{ABC} \phi^A \phi^B \phi^C + \ldots  \Big) \ ,
\ee
where $\phi^A = \Phi^A- \Phi_0^A$ are small fluctuations around the fixed point
value. It is straighforward to continue this expansion if one uses the
fact that for many physical interesting cases $T_{ABC}$ is covariantly
constant, which implies that the scalar manifold is symmetric and
homogeneous \cite{360}.

The mass terms of the scalars are extracted from the expansion of
the potential 
\be504
P = 6 \, W_0^2 ( 1 + {1 \over 3} g_{AB} \phi^A \phi^B -
{32 \over 3} T_{ABC} \phi^A \phi^B \phi^C \pm \ldots  )
\ee
where $6 \, g^2 \, W_0^2$ is the (negative) cosmological
constant and the mass of $\Phi^A$ is
\be506
m_A^2 = 4\, W_0^2 \, \lambda_A \ ,
\ee
with $\lambda_A$ as eigenvalue of $g_{AB}$ calculated at the fixed point.
Following the standard procedure \cite{320} these masses translate into
conformal dimensions of the corresponding operator in the field theory
$\Delta_A = 2\Big(1 + \sqrt{1 + m^2_A/4} \Big) = 2\Big( 1 +
\sqrt{1 + W_0^2 \, \lambda_A}\Big)$. 
In addition to the scalar masses, 
the gauging yields also mass terms for the fermions \cite{020}
\be508
\ba{l}
- i W \bar \psi_{\mu} \Gamma^{\mu\nu}\psi_{\nu}
+ i\, {3 \over 2}\, (g_{AB} W + 4 \sqrt{2} T_{AB}^{\ \ C}
\partial_C W) \, \bar \lambda^A
\lambda^B \ .
\ea
\ee
At the fixed point the gravitino mass term is therefore given by the
cosmological constant and, as required by supersymmetry, the masses of
the gauginos coincide with the scalar masses. Notice, the scalar
masses come from the sugra potential $P$ whereas the fermionic masses
from the superpotential $W$. Let us also mention that the supergravity
potential can vanish identically, $P \equiv 0$, but $W \neq 0$ \cite{370}
(assuming that $T_{ABC}$ is covariantly constant). Obviously, in this
case the scalars remain massless, but the gauginos feel a potential
and supersymmetry is broken, see also \cite{560}.  Furthermore, let us
also mention that all masses are suppressed by the cosmological
constant, which has to be small for a reliable supergravity picture.

Next, following the discussion of \cite{072, 060} we can formulate a
$c$-theorem for the supersymmstric flow. By investigating the Einstein
equations (see eq.\ (24) in \cite{030}) we find
\be410
\ba{rcl}
- R_0^{\ 0} + R_r^{\ r} &=& - 3 g^2 r^2 e^{4U} \Big[U'' + 2U'^2 + 
{3 \over r} U' \Big] = g^2 \Big[  |h|^2 - {2 \over 3} (X\cdot h)^2 \Big] \\
&=&  - 3 g^2 W \mu {d \over d \mu} W 
= g_{AB} \partial \Phi^{A} \cdot \partial \Phi^B  \geq 0 \ .
\ea
\ee
Using the projector $g^{AB} \partial_A X^I \partial_B X^J
= G^{IJ} - {2 \over 3} X^I X^J$ which becomes $g^{AB} \partial_A
W \partial_B W = {1 \over 9} |h|^2 - {2 \over 3} W^2$
this equation can be written as
\be430
W \, \mu {d \over d \mu} W + 3 g^{AB} \partial_A W \partial_B W = 0 \ .
\ee
Hence, $W^2$ is a monotonic decreasing function in $\mu$.  On the
other hand the $c$-theorem conjectures a $C$-function which is
monotonic under the RG flow and becomes the central charge of the CFT at
their extremum. Generically it interpolates between two different CFT with
different central charges.  These central charges, are basically given by
the radius of the AdS space \cite{190}: $c \sim l^3$, up to universal
constants. Recalling that at the fixed point $l = (g|W|)^{-1}$, a natural ansatz
for the $C$-function is
\be420
C(\mu , \Phi^A) \sim {1 \over (g |W|)^3} \ .
\ee
Moreover, expressed in terms of $\mu$ (recall $\mu d\mu = W \, rdr$)
the gaugino variation gives us an expression for the $\beta^A$-function for
the coupling $\Phi^A$
\be440
\beta^A \equiv \mu {d \over d \mu} \Phi^A = g^{AB} \partial_B \log C \ .
\ee
Using the definition of $W$ together with the constraint (\ref{180}) this
represents an exact expression for the $\beta$-function for any
prepotential. Inserting the expansion for $W$ this $\beta$-function
can be expanded as
\be442
\beta^A = - 3 \, g^{AB} {\partial_A W \over W} =
-2\, \phi^A -4 \, T^{A}_{\ BC} \phi^B \phi^C \pm \ldots  \ .
\ee
Using this we can write eq.~(\ref{430}) as the RG equation for the
$C$-function
\be450
\Big( \mu {d \over d\mu} - \beta^A \partial_A \Big) C(\mu , \Phi^B) =0 \ . 
\ee
Since $W^2$ is a strictly decreasing function and for $W>0$, the
$C$-function is strictly increasing under the RG flow
($c$-theorem)
\be452
\mu {d \over d \mu} C \geq 0 \ .
\ee
It also means that the superpotential behaves monotonic during the flow.

Like the entropy in thermodynamics, the $c$-theorem reflects the
irreversibility of a QFT under the RG flow, where the RG parameter $\mu$
represents the scale below which (massive) modes have been integrated
out, for a recent discussion see \cite{200}.  In fact, there seems to be
a close relationship between the gravitational entropy and the central
charge of the field theory. Both quantities can be obtained by an
extremization with respect to the moduli: the central charge from the
$C$-function as described before and the Bekenstein-Hawking entropy from
the BPS mass \cite{220, 230}. As a consequence, both are moduli independent
and depend only on universal quantities like topological data. On the
other hand the BPS mass itself as well as the $C$-function are not
universal, the mass depends on the moduli and the $C$-function is
scheme dependent. To be concrete, the Bekenstein-Hawking entropy of black
holes in 5 dimensions is given by $S= M_{extr.}^{3/2}$, where
$M_{extr.}$ is extremum of the BPS mass as a function of the moduli ($\sim
q_I X^I$) while keeping fix the charges $q_I$.  This definition of the
entropy does not require a horizon or a black hole as it can directly be
derived from the first law of thermodynamics \cite{450}, but see also
\cite{150}. However, this approach has a sublte point - the temperature.
E.g., extremal black holes have a non-vanishing entropy, whereas there is
no natural definition of temperature in an asymptotic flat spacetime. On
the other hand, our situation is different, because anti de
Sitter space has an intrinsic temperature. Black holes, e.g., need a
minimal temperature $T_0 \sim \sqrt{-\Lambda}$ to be in a thermodynamical
equilibrium with the AdS space \cite{520} and at this critical point also
the mass is directly fixed by the cosmological constant $M \sim
1/(-\Lambda)$.  Recalling that the cosmological constant is $\Lambda
\sim W_0^2$, the first law gives for the entropy\footnote{Of course these
arguments  have to be justified by a microscopic or statistical
analysis.} $S = \Big({M \over T_0}\Big)_{extr.} \sim W_0^{-3/2}$, where
the extremum is taken with respect to the moduli while keeping fixed the
parameter $V_I$ (F-I terms). As mentioned in
\cite{150} the extremization with respect to the moduli yields an
extremum for the temperature as well, $T_0$ is therefore a reasonable
candidate for the temperature. Since this extremization is exactly equivalent
to the equation (\ref{350}) calculated at the fixed point
\cite{230}, this analogy suggests that
\be468
c_{ _{FT}} = {1 \over (gW_0)^3} \sim S_{grav}^{-2} \ .
\ee
Note, the field theory central charge decreases in the flow towards the
IR, while the gravitational entropy increases towards the UV and recall,
the gravitational UV regime translates into the field theory IR and vice
versa! One could object, that the equations (\ref{350}) allow in general
for more solutions, but the entropy should be a unique quantity for a
given system. On the other hand, different solutions will correspond to
different fixed scalar fields and therefore translate into different
fixed points and thus there is no puzzle (since the field theory is expected
to be very different at different fixed points).

\begin{figure}
\begin{center}
\includegraphics[angle=-90, width=110mm]{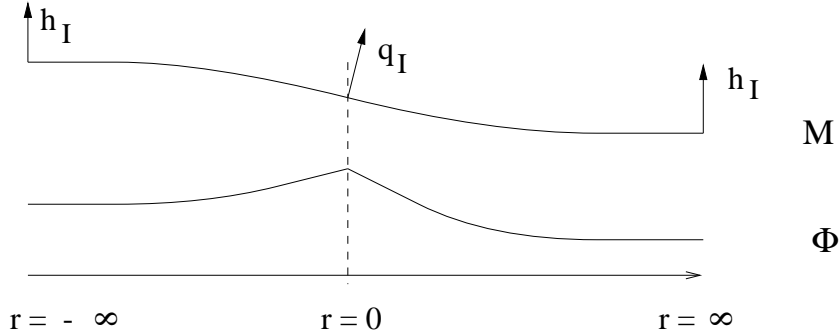} 
\end{center}
\caption{The scalar fields are not constant and therefore every
point on the scalar manifold $M$ represents a different point in
spacetime. A solution of the scalar field equation defines a
trajectory $\Phi = \Phi(r)$ connecting two fixed point
values. Supersymmetry is preserved, if along this trajectory the
normal vector of $M$ remains parallel the harmonic function $H_I = h_I
+ {q_I \over r^2}$ which fixes the solution. Although $M$ is smooth
and differentiable the trajectory may make turns related to source
terms at the domain wall.}
\label{flow}
\end{figure} 

As next step, let us describe the supersymmetric flow in detail.  As
one may have realized, the crucial part of the solution is given by
equation (\ref{350}), which states that the normal vector $X_I$ on the
scalar manifold has to be parallel to $H_I$ or in other words: for a
given radius $r$ a supersymmetric solution is given by the point(s) on
the scalar manifold where $X_I \, \| \, H_I$. Moreover, since $H_I$ is
a vector that interpolates between $h_I$ at $r= \infty$ and $q_I$ near
$r\simeq 0$, every point on a supersymmetric path (flow) corresponds
to a different radius (energy) and along this path the normal vector
{\em has to remain parallel} to the (with $r$ changing) vector
$H_I$. Our solution shows that there is always a path departing from a
fixed point, with the normal vector given by $h_I$ towards any normal
vector $q_I$, see figure \ref{flow}.

But the point $r=0$ {\em is not} a new fixed point. A fixed point would imply
that the metric becomes anti de Sitter, but instead near $r=0$ the
spacetime becomes flat. In order to see this, let us note that as long as
all $q_I$ are non-vanishing, none of the $X_I$ vanishes or
diverges\footnote{To verify this, take arbitrary ratios of eq.\
(\ref{350}).}. Therefore near this point $e^{2U}$ has to scale
like $1/r^2$ and the metric (\ref{200}) becomes flat. Moreover, if it
would be a fixed point, the first term in the gaugino variation had to
vanish, but the first term in (\ref{260}) vanishes only
if $h_I \sim H_I$ (see (\ref{240})). This is the case only if $r\rightarrow
\infty$ or for $h_I \, \| \, q_I$, the latter case would mean that we stay at
a given fixed point. Hence, in order to reach a second fixed point we have
to continue our solution through this point and at $r = - \infty$ we
reach again an anti de Sitter space, i.e., we are again at a
fixed point.

The continuation can be done in a symmetric way, i.e.~by identifying
the solution under $r \rightarrow -r$, but for us more interesting is the
non-symmetric case, where the asymptotic AdS spaces differ. In order to
have a well-defined transition we have to require that the scalars and
the metric join smoothly which is ensured if the $q_I$ vector is the same
on both sides. Also, we do not want to change the gauging while running
from one fixed point to another and therefore we impose that $h_I
\sim V_I$ is the same on both sides. Hence, we use the same set of
harmonic functions on both sides which implies that the normal vector
$X_I$ behaves smoothly at $r=0$. As a consequence of all this, the scalars
and $W$ are regular. In order to ensure that we nevertheless move to {\em
another} fixed point and not to come back to our starting point, we have to
change the topology of the internal manifold at $r=0$. There has to be a
phase transition, which changes the intersection form (\ref{180}). An
example is a flop transition, which
implies that the intersection form gets an additional term ${\cal V}
\rightarrow {\cal V} - {1 \over 6} t_2^3$ (if we assume that the cycle
parameterized by $t_2$ vanishes at the transition point). For this type
of transition, not only the scalar manifold but also the first and
second order derivatives pass the transition point at $t_2 = 0$ smoothly,
but higher derivatives, as typical for phase transitions, may jump;
see also the discussion in \cite{090}. As we have shown earlier, the
second derivative of $W$ is positive definite at the fixed points (scalar
mass terms), suggesting that $W$ has only minima and no saddle points and
extrema (for $W>0$). But the information about the first and second order
derivative is not enough to get a global picture of $W$, instead
we have to take into account that the scalar manifold (moduli space) has
boundaries and typically at these boundaries massless states appear and
second order derivatives of $W$ vanishes, see $W_{+}$ in (\ref{620}). For
a strictly non-degenerate K\"ahler metric, the number of minima should be
related to the number of boundaries and if there is only one boundary
like for torus compactification (see figure 2 in \cite{020}) we get an
unique extremum for $W$.

The smoothness of the scalar manifold does not mean that the
trajectory of the flow given by $\Phi = \Phi(r)$ is differentiable at
$r=0$. In the symmetric case where we identify both sides it is rather
like a reflection, where the scalars have maximal velocity at the
wall.  At this point the velocity ($\sim \partial_r \Phi$) changes its
sign, which produces a $\delta$-function in the second derivative and
indicates source terms located at the wall \cite{390}. In the
non-symmetric case, there is still a sign change in the velocity, but
it is rather a non-complete reflection or a refraction; see also the
discussion in \cite{530, 180}. Nevertheless, the underlying manifold
should be smooth, only the flow trajectory is making a turn at
$r=0$. Let us also mention, that the vector $q_I$ determines the
trajectory or equivalently the $q$'s are allowed deformations of a
given trajectory. But not for all choices of $q_I$ we reach a phase
transition point, i.e., for those cases the flow has to come back to
the starting point.

Also, one may ask whether the superpotential has more extrema for
a given model with a given intersection form (recall, that we
still assume trivial hyper multiplets). Any extremum of the
superpotential implies an AdS vacuum and allows for some unbroken
supersymmetries and therefore, has to be part of our solution. As
we discussed earlier, supersymmetry requires that the normal vector of
the scalar manifold (i.e., $X_I$) becomes at the fixed point parallel to
the {\em constant} vector $h_I$ or $V_I$, i.e., $h_I
\partial_A X^I = 0$.  Therefore, extrema of the superpotential coincide 
with extrema of the scalar manifold with respect $h_I$, see figure
\ref{flow}. However, the scalar manifold as defined by the
cubic equation ${\cal V} = 1$, is generically not connected and the
different branches are separated by singular lines where the
prepotential vanishes ${\cal V} =0$. For many physical interesting
cases this manifold is a symmetric non-compact space with constant
curvature \cite{240, 020}, which implies that every branch allows for
at most one extremum, i.e., a point where the normal vector is
parallel to $h_I$ or $h_I \partial_A X^I = 0$.  If a given
prepotential allows for multiple solutions inside a given K\"ahler
cone, they have to lie on different branches, i.e., they should be
disconnected.  No flow is possible between two extrema inside a given
K\"ahler cone, one has to pass a phase transition to reach a second
extremum of the superpotential! For the field theory this means, that
between two fixed points of the $\beta$-function there has to be a phase
transition in the supergravity description.

\begin{figure}
\begin{center}
\includegraphics[angle=-90, width=130mm]{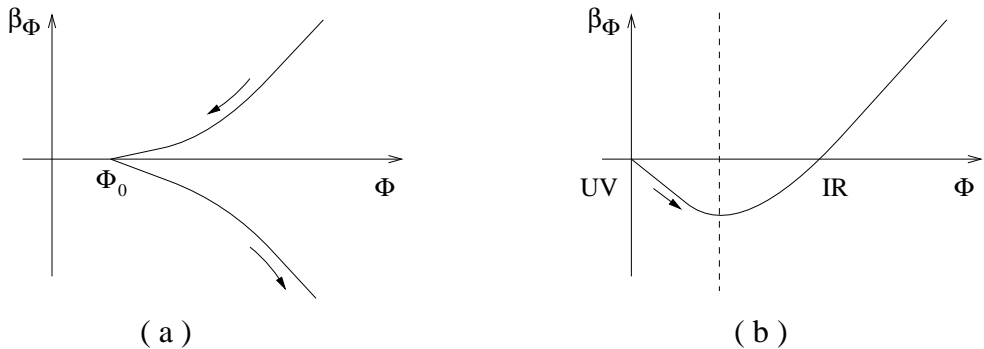} 
\end{center}
\caption{Case (a) shows a coupling that runs towards a non-vanishing
IR of UV fixed point value $\Phi_0$ (the arrow indicates the IR flow).
On the other side, case (b) shows a running coupling with two
fixed points. This case is typical for N=1 super Yang-Mills model.
}
\label{RG-flow} 
\end{figure}


\section{Examples}


Let us start with a simple example that can be solved completely.
It is given by the prepotential
\be700
{\cal V} = X^0 X^A \eta_{AB} X^B
\ee
and corresponds in the ungauged case to a compactification on $K3 \times T_2$
with an appropriate signature for $\eta_{AB}$. Therefore it will give us
a supergravity picture of branes at $Z_k$ orbifold singularities, as
e.g., described in \cite{470}. First we have to solve the equations
(\ref{350}), which become
\be710
H_0 = e^{2U} X^A \eta_{AB} X^B \qquad , \qquad H_A = 2 \, X^0e^{2U} \,
\eta_{AB} X^B \ .
\ee
Fixing $X^0$ by the requirement ${\cal V}=1$ and taking $X^A = \Phi^A$ as
physical scalars we find as solution
\be720
X^A = e^{-4U} \, \eta^{AB} H_B H_0 \quad , \quad e^{3U} = {1 \over 2}
\sqrt{ H_0 H_A \eta^{AB} H_B} \ .
\ee
The superpotential becomes
\be730
W = {V_0 \over \Phi_A \Phi^A} + V_A \Phi^A
\ee
which allows for a minimum ($\partial_A W = 0$) where
\be740
W_0^{3/2} = {1\over 2} \sqrt{ 27 \, V_0 (V_A \eta^{AB} V_B)}
\quad , \quad \Phi^A = 2^{1/3} { V_0 V_A \over (V_0 V_A \eta^{AB}
V_B)^{2/3}} 
\ee
and $g^2 W_0^2 = - \Lambda$ is the cosmological constant which defines
also the central charge of the CFT. We can also calculate the
$\beta$-function as a function of the scalars and find (see
eq.~(\ref{442}))
\be750
\ba{rcl}
\beta^A &=& -3 g^{AB} {\partial_B W \over W} \\
&=& -  {1 \over V_0 + (\Phi \cdot \Phi)(\Phi \cdot V)}
\Big[4 \Phi^A (\Phi \cdot \Phi)(\Phi \cdot V) -  2 V_0 \Phi^A - 3 
(\Phi \cdot \Phi)^2 V^A \Big] \ .
\ea
\ee

But not only the supergravity side, also the field theory can be explored
more explicitly. E.g., this model provides a sugra picture for branes at
$Z_k$ orbifolds as described in \cite{470}, where a subclass of our
scalars corresponds to blow-up modes of 2-cycles and the charge parameter
$q_I$ entering the harmonic functions corresponds to the number of branes
which are on top of each other. As consequence of the orbifold, the gauge
group factories and the sugra scalars (moduli) parameterize the space of
gauge couplings for the different $U(q_I)$ gauge groups. The fixed point
values of these gauge couplings are inversely related to the fixed
scalars in (\ref{740}) and the central charge is $c \sim W_0^{-3}$. The
non-vanishing $\beta$-function means the conformal symmetry is
generically broken, but for special values of $V_I$ the sugra scalars are
constant everywhere. This is exactly the case if $q_I \| V_I$ where the
radial dependence of the scalars drops out (no running) and the domain
wall becomes exactly $AdS_5$. For black holes, this case is known as
double extreme solution (an extremal black hole with extremal mass).

As a second example we will discuss a topological non-trivial case, which
has two fixed points.
The dashed line indicates the supergravity phase transition at $r=0$
between the distinct vacua and in our example we consider a transition
related to an additional cubic term in the intersection form,
as it  has been discussed for black holes with constant scalars in
\cite{090} and later on for non-trivial scalar in \cite{100}. The two
phases correspond to different triangulations of an elliptically
fibered Calabi-Yau with the base $F_1$; for more information about
this flop transition we refer to \cite{210, 080} and for a general
discussion about Calabi-Yau phase transitions to \cite{170}. For this
Calabi-Yau exists an orientifold limit \cite{510} and the type I dual
description (compactified on K3) has been discussed in \cite{500}. We
expect that in the dual field theory the space of gauge couplings can
again be parameterized by our supergravity scalars\footnote{Notice, on
the supergravity side the dilaton sits in the universal hypermultiplet
which is fixed in our setup (not running) and therefore also on the
field theory side we have to keep fix the over-all norm of the gauge
couplings as well. In this respect it would be interesting to explore
the domain wall solution of \cite{260} which has a non-trivial
dilaton.}.

Let us come to the concrete model. One phase, which corresponds to the
second Chern class $c_2 = (92 , 102, 36)$ is described by the
prepotential \cite{080}
\be510
{\cal V}_{+} = {3 \over 8} (X^2)^3 + {1 \over 2} X^2(X^1)^2 - {1 \over 6} 
(X^3)^3 
\ee
and as it will turn out at the end, this model yields the bigger central
charge.
The coordinates that we are using are related to the Calabi-Yau
moduli by 
\be520
t_1 = X^3 \quad , \quad t_2 = X^2 - X^3 \quad , \quad 
t_3 =X^1-{3 \over 2} X^2 \ .
\ee
In order to stay inside the K\"ahler cone all $t$'s have to be positive
or equivalently: ${2 \over 3}X^1 > X^2> X^3> 0$. We reach the boundaries
of the moduli space at $t_1=X^3=0$ (elliptic fibration over $P_2$) and at
$t_3= X^1 -{3 \over 2} X^2 = 0$ where tensionless strings emerge. As a
side remark, these $t$-moduli correspond to different 2-cycles and {from}
the F-theory perspective our domain wall can be seen as 7-branes with four
coordinates wrapped in the internal space. Then, the tensionless strings
can either be understood as 3-branes wrapping a vanishing 2-cycle which
is not part of the 7-brane, or as zero-size instantons on the world
volume, see \cite{210} for more details. Let us recall, we
consider gauged supergravity which, as mentioned before, corresponds to
non-trivial (non-zero mode) fluxes for some cycles. Continuing our
discussion, we pass the flop transition at
\be522
t_2 = X^2-X^3 = 0 \ .
\ee
After this transition we enter a different CY with $c_2 = (92, 36
,24)$, which corresponds to the prepotential \cite{080}
\be530
{\cal V}_{-} = { 5 \over 24} (X^2)^3 + {1 \over 2} X^2 (X^1)^2 - {1 \over 2}
X^2 (X^3)^2 + {1 \over 2} (X^2)^2 X^3
\ee
where the Calabi-Yau coordinates are now given by
\be540
\tilde t_1 =X^2 \quad , \quad \tilde t_2 = X^3-X^2 \quad , \quad 
\tilde t_3 = X^1 - {1\over 2} X^2 - X^3 \ .
\ee
The K\"ahler cone is again defined by the domain of positive $\tilde t$'s;
the moduli space ends at $\tilde t_1 =0$ (again tensionless strings emerge)
and at $\tilde t_3 =0$ where an $SU(2)$ symmetry enhancement occurs.  The
flop transition is at $\tilde t_2 =0$.  

Let us discuss the two phases separately.

{\bf 1) The ``+'' phase.} In order to find the scalars we have to solve
(\ref{350}) which gives for the prepotential ${\cal V}_{+}$
\be552
\ba{l}
H_1 = e^{2U} X^1 X^2 \ , \\
H_2 = e^{2U} \Big[{9 \over 8} (X^2)^2 + {1 \over 2} (X^1)^2 \Big] \ , \\
H_3 = e^{2U} \Big[ - {1 \over 2} (X^3)^2\Big] \ .
\ea
\ee
These equations have more solutions, but only one fulfills the conditions
${2 \over 3} X^1 > X^2 > X^3 >0$ (i.e., lies inside the K\"ahler cone)
\cite{090}, \cite{100}
\be550
\ba{l}
X^1 = e^{-U} \sqrt{ H_2 + \sqrt{H_2^2 - { 9 \over 4} H_1^2}} \ , \\
X^2 = {2\over 3} e^{-U} \sqrt{ H_2 - \sqrt{H_2^2 - { 9 \over 4} H_1^2}}\ , \\
X^3 = e^{-U} \sqrt{ -2 H_3} \ , \\
e^{2U} = {1 \over 3} \Big[ H_1 X^1  + H_2 X^2 + H_3 X^3 \Big] 
\ea
\ee
where the harmonic functions have to satisfy
\be560
H_2 \geq {3 \over 2} H_1 \quad , \quad 
H_2 + \sqrt{H_2^2 - { 9 \over 4} H_1^2} \ \geq \ {9 \over 2} |H_3| \ .
\ee
At the transition point where $t_2 = \tilde t_2 = 0$ we have to ensure
that $X^2=X^3$, which translates into a condition for the charges
\be562
{2 \over 9} \Bigg( \ q_2 - \sqrt{ q_2^2 - { 9 \over 4} q_1^2} \ \Bigg)
= |q_3| \ ,
\ee
(recall: $q_3 < 0$). Recall, the $q$'s are deformation parameter
for the flow.

{\bf 2) The ``-'' phase.} Here we have to consider the prepotential ${\cal
V}_{-}$ and find
\be570
\ba{l}
H_1 = e^{2U} X^2 X^1 \ , \\
H_2 = e^{2U} \Big[{5 \over 8} (X^2)^2 + {1 \over 2} 
(X^1)^2 - {1 \over 2} (X^3)^2 + X^2 X^3 \Big] \ , \\
H_3 = e^{2U} \Big[ - X^2 X^3 + { 1\over 2} (X^2)^2 \Big] \ .
\ea
\ee

Again these equations have more solutions, but only one of them can be
connected at $r=0$ to the solution (\ref{550}).  It is given by \cite{090},
\cite{100}
\be580
\ba{l}
X^1 = e^{-U} {\sqrt{2} H_1 \over \sqrt{H_2 + {1 \over 2} H_3 +
\sqrt{ (H_2 + {1 \over 2} H_3 )^2 - 2(H_1^2 - H_3^2)}}} \ , \\
X^2 = {1 \over \sqrt{2}} e^{-U} \sqrt{ H_2 + {1 \over 2} H_3 +
\sqrt{ (H_2 + {1 \over 2} H_3 )^2 - 2(H_1^2 - H_3^2)}} \ , \\
X^3 = \Big({1 \over 2} X^2 - { H_3 \over X^2} e^{-2U} \Big) \\
e^{2U} = {1 \over 3} \Big[ H_1 X^1  + H_2 X^2 + H_3 X^3 \Big] \ .
\ea
\ee
Notice, at the transition point we did not change the harmonic functions,
neither $q_I$ nor $h_I$, we changed only the prepotential.

In order to simplify the situation further, we can go to the symmetry
enhancement line, which means for the ``+'' phase: $X^1 = {3 \over 2} X^2$ 
and for the ``-'' phase: $X^3 = X^1 - {1 \over 2} X^2$. In both cases 
this translates into one condition for the harmonic functions
\be590
H_2 = {3 \over 2} H_1 \ .
\ee 
After imposing this constraint, we have only one physical scalar and we
can describe the situation as shown in figure \ref{RG-flow} (b).  For
this we regard $X^3 = \Phi$ as the physical scalar and we start our flow
in the ``+'' phase at the boundary of the $X^3$ 
modulus, i.e., at $X^3=t^3=0$. It means, we have to set $h_3 = 0$ and find
\be600
\Phi \rightarrow \left\{ \ba{ll} 0 \ ,  & {\rm in \ the \ ``+'' phase}\ 
(r \rightarrow + \infty) 
\\ (48/5)^{1/3}\ ,  &{\rm in \ the \   ``-'' phase} \ 
(r \rightarrow - \infty) \ .
 \ea \right.
\ee
The superpotentials read ($V_1 = 2/3 \, V_2 = V$; $V_3 =0$)
\be620
\ba{l}
W_{+} = V \Big({72 \over 7}\Big)^{1/3} \Big[ 1 + {1 \over 6} \Phi^3
\Big]^{1/3} \ , \\
W_{-} = V \Big[ {5 \over 3} \, \sqrt{f(\Phi)} + {1 \over f(\Phi)}\Big] \ ,
\ea
\ee
where $f(\Phi) = \sqrt{\sqrt{ 9/4 \, \Phi^2 + 3} \; - \; 3/2\, \Phi}$.
Obviously, the extremum in the ``+'' phase is at $\Phi=0$ and the second
derivative of $W$ vanishes at the extremum, which means that the K\"ahler
metric degenerates. This confirms our expectations, because this extrema
lies on the boundary of the moduli space where tensionless strings
appear. In the ``-'' phase the situation is 
different, the fixed scalars as well
as the second derivative is non-vanishing. The ratio of the extrema
gives the ratio of the central charges 
\be610
{c_{-} \over c_{+}} = \left({W_{+} \over W_{-}}\right)_{extr.}^3 = 
{(e^{6U})_{+} \over (e^{6U})_{-}} = {24 \over 25} \ .
\ee


\section{Conclusion}


In this paper we employed domain wall solutions to give a supergravity
description of the RG flow of 4-d super Yang-Mills. On the
supergravity side the domain wall interpolates between two asymptotic
AdS spaces on both sides and is a solution of 5-d gauged supergravity,
where a $U(1)$ group of the R-symmetry has been gauged. In this setup
the running of the supergravity scalars between two fixed point values
corresponds to the running of couplings of operators in super
Yang-Mills. These fixed point values are reached in the two asymptotic
AdS spaces on both sides of the domain wall. Identifying both sides
yields an $S/Z_2$ orbifold in the radial direction, but it does not
describe a flow between {\em different} fixed points. Instead, to reach a
second fixed point one has to assume that a phase transition takes place
while passing the domain wall. The natural framework to discuss this
transition is N=2 supergravity and we have discussed an explicit
example.
This phase transition happens at a boundary of the K\"ahler cone and
therefore, while passing the domain wall in spacetime one leaves a
given K\"ahler cone as well.

Exploring the equations of motion and the supersymmetry variations we
discuss the $c$-theorem and give a supergravity expression for the
field theory $\beta$-functions. These expressions are non-perturbative
in the sense that they hold for any prepotential of N=2 supergravity
but they are restricted to the case of trivial
hypermultiplets. Expanding the supergravity potential as well as the
superpotential around a given fixed point we obtain the cosmological
constant and the masses for the scalar fields. The cosmolgical
constant, which suppresses all mass terms, is obtained by an
extremization of the superpotential with respect to the moduli which
suggests a relation to the  5-d gravitational entropy.

At the end we discussed two examples in more detail. One is a gauging of
the K3 compactification of IIB string theory which should be dual N=2
super Yang-Mills. As expected this has only one fixed point. In the
second example we discuss a gauged model of elliptically fibered
Calabi-Yau with $F_1$ basis, which allows for an orientifold limit and is
dual to type I on K3. This second example exhibits a phase transition and
passing this transition corresponds to a running coupling between
different fixed points of the dual N=1 super Yang Mills. In both cases
the field theory gauge group is expected to be a direct product of
different $U(q_I)$, at least in the orbifold/orientifold limit, and we
argue that deformations of the relative gauge couplings are parameterized
by the supergravity scalars.

There are a couple of interesting directions that are worthwhile to
explore. One is to take into account a non-trivial dilaton, e.g., by
exploiting the solution given in \cite{260}. A further direction is to
investigate the electric-magnetic duality. In general by gauging we break
the duality and therefore we cannot expect dyonic solutions, but it may
happen that two different models flow to the same fixed point and at this
fixed point they are dual to each other. In fact for non-supersymmetric
vacua (with vanishing potential) generalized electro-magnetic dualities
have been discussed in \cite{310}. Of course, to investigate
non-supersymmetric vacua, would be interesting in its own. Finally, let
us also mention that having the explicit supergravity solutions, it is
straightforward to calculate the effective supergravity action. As they
solve the equations of motion only surface terms will survive, which
should generate the amplitutes of super Yang Mills.


\bigskip

{\bf Acknowledgements}

\medskip

I am grateful to Mirjam Cvetic, Oren Bergman and Eric Gimon for discussions.
The work is supportet by a Heisenberg Fellowship of the DFG
and in part by the Department of Energy under grant number
DE-FG03-92-ER 40701.


%
%

\providecommand{\href}[2]{#2}\begingroup\raggedright\endgroup


\end{document}